\documentclass[reprint,superscriptaddress,nofootinbib,amsmath,amssymb, aps,prl,onecolumn,notitlepage]{revtex4-1}

\usepackage[a4paper, margin=2.50cm]{geometry}

\usepackage{bbm}
\usepackage{graphicx}
\usepackage{fancyhdr, color}
\usepackage{hyperref} 
\usepackage{pifont}
\usepackage{float}

\begin{document}

\newcommand{\chqfluct}{\cite{Funo2018}} 
\newcommand{\chqtraj}{\cite{Elouard2018}} 
\newcommand{\chinfera}{\cite{Croucher2018}} 
\newcommand{\chthinf}{\cite{Bera2018}} 
\newcommand{\chsebox}{\cite{Koski2018}} 

\title{Maxwell's demon in superconducting circuits}

\author{Nathana\"el Cottet}
\affiliation{Laboratoire Pierre Aigrain, \'Ecole normale sup\'erieure, PSL Research University, CNRS, Universit\'e Pierre et Marie Curie, Sorbonne Universits\'e, Universit\'e Paris Diderot, Sorbonne Paris-Cit\'e, 24 rue Lhomond, 75231 Paris Cedex 05, France}
\affiliation{Universit\'e Lyon, ENS de Lyon, Universit\'e Claude Bernard Lyon 1, CNRS, Laboratoire de Physique, F-69342
Lyon, France}

\author{Benjamin Huard}
\email{benjamin.huard@ens-lyon.fr} 
\affiliation{Universit\'e Lyon, ENS de Lyon, Universit\'e Claude Bernard Lyon 1, CNRS, Laboratoire de Physique, F-69342
Lyon, France}

\date{\today}

\begin{abstract}
This chapter provides an overview of the first experimental realizations of quantum-mechanical Maxwell's demons based on superconducting circuits. The principal results of these experiments are recalled and put into context. We highlight the versatility offered by superconducting circuits for studying quantum thermodynamics.
\end{abstract}

\maketitle

\thispagestyle{fancy}

\section{INTRODUCTION} \label{sec:intro}
The past decades have seen the development of superconducting circuits based on Josephson junctions as one of the most promising platforms for  quantum information processing~\cite{Devoret2013a}. Owing to their high level of control in both their design and their manipulation, they naturally constitute a convenient testbed of fundamental properties of quantum mechanics. Superconducting circuits reach strong coupling with microwave light, allowing quantum-limited amplification~\cite{Roy2018}, strong Quantum Non Demolition measurement~\cite{Mallet2009}, weak measurement~\cite{Murch2013a,Hatridge2013}, quantum feedback~\cite{Vijay2012c}, and the observation of quantum trajectories~\cite{Weber2016}. From a quantum thermodynamics point of view, this high level of control gives full access to the dynamics of energy and entropy flows between the different parts of the experimental system. Up to now, three experimental realizations of a  Maxwell's demon have been achieved using superconducting circuits in the quantum regime~\cite{Masuyama2017,Naghiloo2018,Cottet2017}. They all consist of a 3D-transmon qubit dispersively coupled to a 3D cavity waveguide measured at cryogenic temperatures around 20 mK~\cite{Paik2011a}. The characteristic frequencies of such systems are in the microwave range.
\paragraph{} Szilard reformulated the original Maxwell's demon gedanken experiment in the case of a single molecule in a two sided box~\cite{Szilard1964,Maruyama2009}. In general, one can cast the experiment in terms of five components with different roles: system, demon, two thermal baths and battery. At the beginning of each thermodynamic cycle, the system is thermalized to its thermal bath. The demon then acquires information on the system to extract work, which can then charge a battery. The apparent possibility to extract work out of a single heat bath vanishes when considering the need to reset the demon state in order to close the thermodynamic cycle~\cite{Landauer1961,Bennett1982a}. One way to reset the demon state consists in thermalizing it with a hidden thermal bath at the end of the cycle, or by actively resetting its state at the expense of external work. There are plenty of ways to transfer information, extract work and thermalize the system. From an experimental perspective, the manner work and entropy flows are measured or inferred is also crucial since the measurement of a quantum system is inherently invasive. This chapter will therefore focus on the existing experimental realizations that illustrate what superconducting circuits can bring to quantum thermodynamics. The chapter is organized as follows. We first introduce the reader to the field of circuit Quantum Electro-Dynamics. Then we present the spirit of the three existing experiments before describing the particular experimental realizations in details.
\subsection{Introduction to circuit-QED}

In this section, we provide a brief introduction to circuit-QED. The interested reader is advised to look into a recent review on the subject~\cite{Wendin2016,Gu2017}.
\paragraph{} A superconducting qubit that is coupled to a cavity can reach two main regimes of interest. First, close to resonance, they can swap excitations, which results in vacuum Rabi splitting. In this chapter, we focus on the opposite regime, where the cavity-qubit detuning is much larger than their coupling rate. This so called dispersive regime can be described by the Hamiltonian~\cite{Blais2004}
\begin{equation} 
H_\mathrm{disp}=\frac{\hbar \omega_q}{2}\sigma_z+\hbar \omega_ca^\dagger a-\hbar\frac{\chi}{2} a^\dagger a\sigma_z\ ,
\label{H}
\end{equation}
where $a$ is the annihilation operator of a photon in the cavity, $\omega_q$ (respectively $\omega_c$) the frequency of the qubit (resp. cavity), and $\sigma_z$ the Pauli matrix of the qubit along $z$. The two first terms represent the Hamiltonians of the qubit and cavity, while the last term describes the coupling between them. Compared to the case of the ground state of both qubit and cavity, the interaction induces a frequency shift $-\chi$ of the cavity when the qubit is excited while the qubit frequency is shifted by $-N\chi$ when the cavity hosts $N$ photons. Thanks to this coupling term it is possible to entangle the qubit and the cavity and hence to transfer information between the two. Moreover, when the cavity is coupled to a transmission line, this information can be either dissipated in the environment or collected into a measurement apparatus.
\paragraph{}The state of the qubit and cavity is controlled using microwave drives on or near resonance with either the qubit or the cavity. Let us consider first a drive near qubit frequency at $\omega_q-\delta$. Without loss of generality one can set the phase of the drive so that it is along the $y$-axis of the Bloch sphere. In the rotating frame of the drive and only keeping the slowly rotating terms (rotating wave approximation) the Hamiltonian becomes
\begin{equation}
H^\mathrm{q}_\mathrm{driven}=\frac{\hbar}{2}(\delta-\chi a^\dagger a)\sigma_z+\Omega_q\sigma_y\ ,
\label{Hq}
\end{equation}
where $\Omega_q$ is proportional to the amplitude of the drive. This Hamiltonian induces Rabi oscillations of the qubit around an axis, which depends on the number of photons in the cavity. Energetically the qubit undergoes cycles of absorption where work is absorbed from the drive and stimulated emission where work is emitted in the drive. Similarly a drive near cavity frequency at $\omega_c-\Delta$ gives the following Hamiltonian
\begin{equation}
H^\mathrm{c}_\mathrm{driven}=\hbar(\Delta-\frac{\chi}{2}\sigma_z)a^\dagger a+\Omega_c(a+a^\dagger)\ ,
\label{Hc}
\end{equation}
where the complex drive amplitude, proportional to $\Omega_c$, is here chosen to be positive. The result is a displacement of the cavity field that depends on the state of the qubit. Assuming the cavity is initially in vacuum it results in the preparation of a coherent state $|\alpha_g\rangle$ (respectively $|\alpha_e\rangle$) in the cavity when the qubit is in the ground (respectively excited) state. Note that two coherent states are never fully orthogonal ($\langle\alpha_e|\alpha_g\rangle\neq 0$) so that they cannot perfectly encode the qubit state.
\paragraph{}All the processes described so far are unitary. In the Zurek description of a quantum measurement~\cite{Zurek2003}, driving the cavity corresponds to a \textit{premeasurement} of the qubit state. The information stored in the cavity eventually escapes towards the transmission line, and can therefore be amplified and detected by classical detectors hence terminating the measurement process of the qubit's state. In the dispersive regime, the observable $\sigma_z$ commutes with the qubit-cavity Hamiltonian~(\ref{H}), ensuring that the measurement is Quantum Non Demolition.

\subsection{Description of the existing experiments}
All three experimental realizations presented in this chapter share the common feature of using the qubit as the system. Its state is measured by the demon (of different nature depending on the experiments) thanks to the coupling term of the dispersive Hamiltonian~(\ref{H}). Work is extracted through a pulse on resonance with the qubit that induces a rotation of the qubit. The pulse acts as the battery~\footnote{Strictly speaking, the battery is the propagating electromagnetic mode that contains the pulse. It can both store and use the energy it contains, hence qualifying as a battery. Indeed, it can store the excitation of a qubit or of a classical cavity field as described in the text. Moreover, if it interacts with an other distant ancillary qubit or cavity in its ground state, it can provide work to excite it. In the text, the qubit extracted work is used to amplify the pulse in the battery.} and is powered-up when the qubit is flipped from a high-energy state to a lower-energy one.

\begin{figure}[H]
\begin{center}
\includegraphics[width = 0.4 \textwidth]{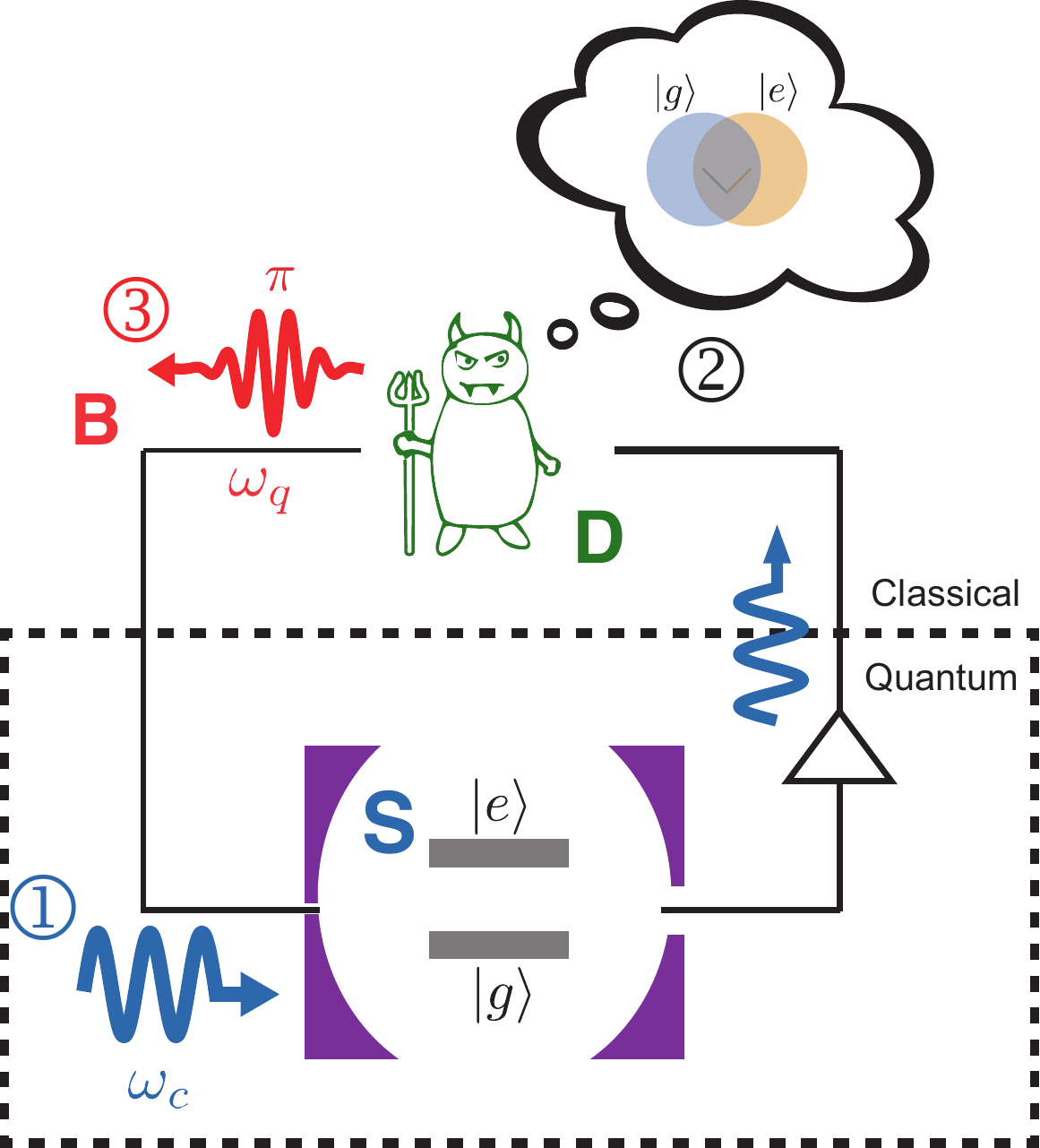}
\caption{\label{masuyama} Demon whose action is based on a single measurement. \ding{172} After qubit initialization, a pulse at cavity frequency is transmitted through the cavity so that its phase encodes the state of the qubit. \ding{173} This information is recorded by a classical measurement apparatus acting as the demon. \ding{174} A feedback $\pi$-pulse is applied conditioned on the measurement outcome in order to extract work.}
\end{center}
\end{figure}

\paragraph{}Masuyama \emph{et al.}~\cite{Masuyama2017} base their Maxwell's demon on a measurement-based feedback scheme. After initialization, the qubit is measured and feedback control is used conditionally on the result of the measurement in the following way: whenever the qubit is measured in $|e\rangle$, a $\pi$-pulse flips it back to the ground state and transfers one quantum of work to the battery. In contrast when the qubit is measured in $|g\rangle$ no pulse is applied. The operation time of the sequence is much faster than the thermalization time of the qubit with the rest of the environment so that the whole process can be considered adiabatic. In this experiment the demon is therefore the classical measurement apparatus and information is acquired and stored into a classical memory (Fig.~\ref{masuyama}). An interesting twist is added by the possibility to use a weak measurement for the feedback control input. Masuyama~\emph{et al.} are then able to demonstrate the role of mutual information in the second law for quantum systems.

\begin{figure}[H]
	\begin{center}
	\includegraphics[width = 0.4 \textwidth]{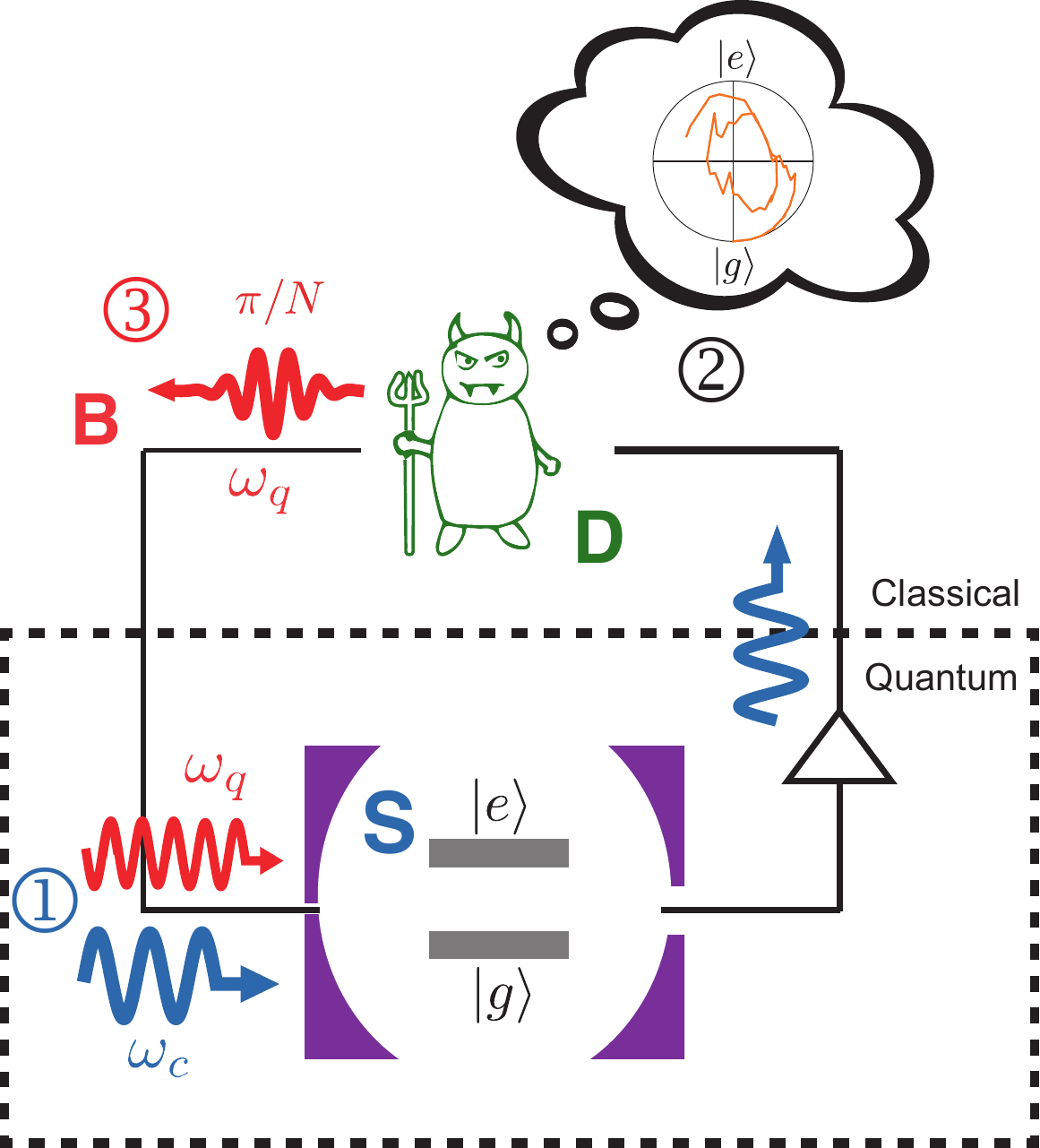}
\caption{\label{naghiloo}Trajectory based demon. \ding{172} After initialization, the qubit is driven at $\omega_q$ while a weak tone at $\omega_c$ measures its state. \ding{173} The information is recorded in a time-resolved way, allowing the demon to reconstruct the quantum trajectory of the qubit in the $XZ$ plane of the Bloch sphere. \ding{174} Based on this information, an optimized feedback pulse is applied to flip down the qubit to the ground state and extract work.}
\end{center}
\end{figure}

\begin{figure}[H]
	\begin{center}
	\includegraphics[width = 0.4 \textwidth]{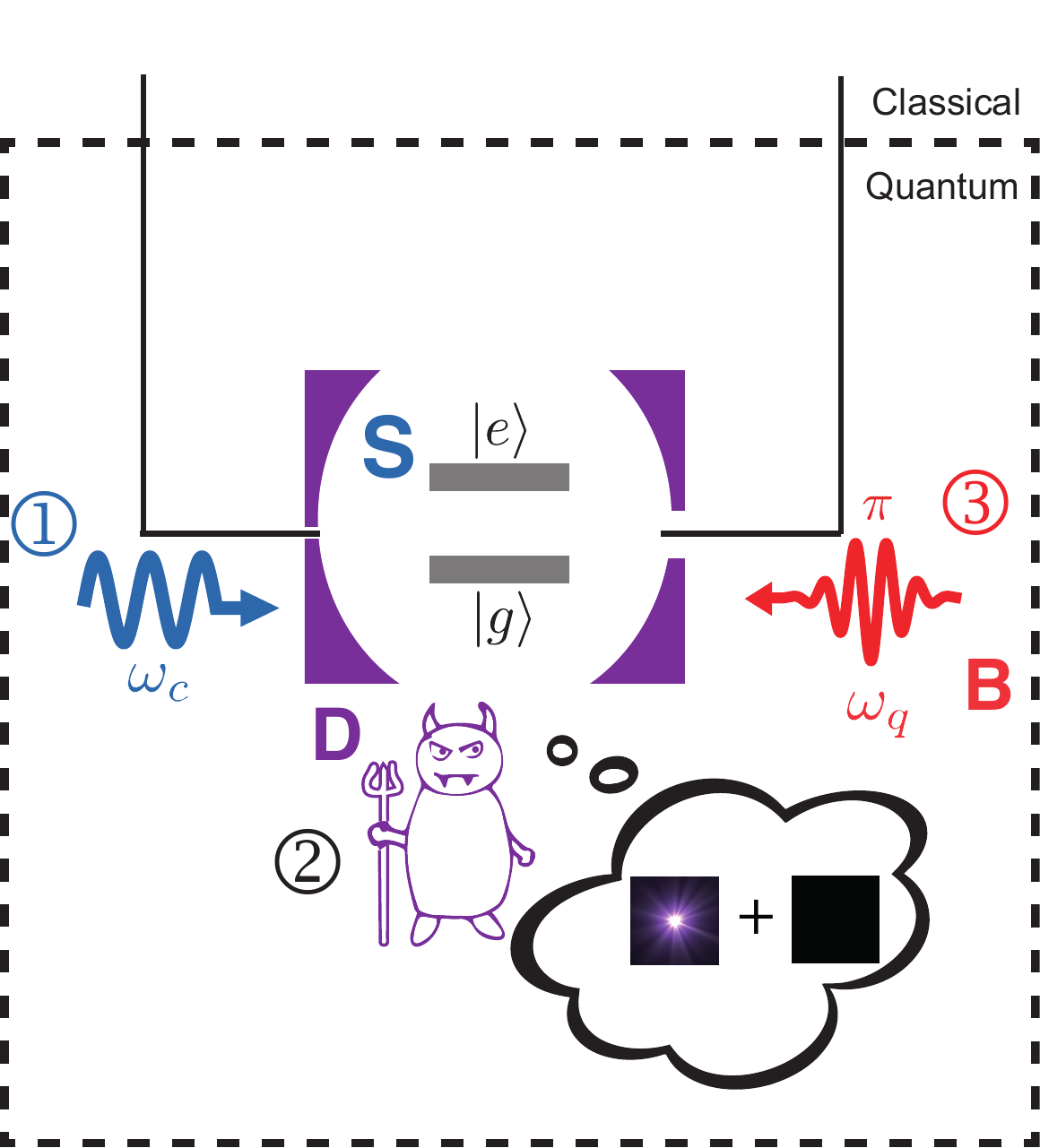}
\caption{\label{cottet}Autonomous quantum demon. \ding{172} After initialization the cavity is populated conditioned to the qubit being in $|g\rangle$ using a drive at $\omega_c$. \ding{173} The cavity state reflects the qubit state, hence acting as the demon and possibly exhibiting quantum coherences. \ding{174} A $\pi$-pulse at $\omega_q$ extracts work conditioned to the cavity being in the vacuum state. Importantly, the information never leaves the quantum world during the whole process.}
\end{center}
\end{figure}

\paragraph{}Naghiloo \emph{et al.}~\cite{Naghiloo2018} also present a Maxwell's demon based on a classical detector. In their case the demon tracks the quantum trajectory of the qubit thanks to time-resolved  measurement records. In this case, after initialization, the qubit is driven on resonance while a weak measurement tone is applied at cavity frequency. The qubit state is then reconstructed using the measurement records based on the stochastic master equation (see Appendix). This classical detector acts as a demon that uses its knowledge on both the qubit excitation and coherences to apply an optimal feedback pulse that flips the qubit to the ground state hence extracting work out of the qubit (Fig.~\ref{naghiloo}). Importantly in this experiment, the qubit exchanges work with the qubit drive during the measurement process. Using the quantum trajectory, one can determine how much work is exchanged at each time step. Interestingly, this amount of work cannot be controlled and present a stochastic behavior (see chapter~\chqtraj). This experiment confirms the crucial role of mutual information in the second law for quantum systems.
\paragraph{}In the previous experiments, the demon is a classical black box. The resolution of the paradox of the Maxwell demon involves the acknowledgment of the demon's information as a physical object. In order to analyze the inner dynamics of the demon and even probe its quantum coherence, Cottet, Jezouin \emph{et al.}~\cite{Cottet2017} demonstrated an autonomous Maxwell's demon in the quantum regime (classical autonomous demons using single electron transistors are discussed in chapter~\chsebox). After initialization in a thermal or a superposed state, a pulse at $\omega_c$ is applied on the cavity and displaces it conditioned on the qubit being in the ground state. It is followed by a $\pi$-pulse at $\omega_q$ that flips the qubit conditioned to the cavity hosting 0 photon. This sequence is realized in a time smaller than the lifetimes of both the qubit and cavity so that the information stored in the cavity does not have the time to escape into the transmission line. Therefore the demon is here the cavity whose quantum state could be measured in a quantum state that exhibits quantum coherences (Fig.~\ref{cottet}). Another particularity of this experiment is the direct measurement of the work extracted into the battery. The other experiments use a Two Point Measurement protocol, which is described below.

\section{QUANTUM-CLASSICAL DEMON}

\subsection{Inferring work and tuning the measurement strength}

Before detailing how fluctuation relations can be investigated using superconducting circuit based quantum-classical demons, we discuss two key tools for the realization of these experiments.

\subsubsection{Inferring work from Two Point Measurement}
Acquiring information on a quantum system is known to be  invasive: if the quantum system is not in an eigenstate of the measured observable, the outcome of the measurement is non deterministic and the system state changes following measurement. Work is not an observable~\cite{Talkner2007}. Therefore quantifying the work done on a quantum system is subject to interpretation. However, there is one case that does not suffer from these inconsistencies. It is the work done on a system that starts from an eigenstate of the Hamiltonian, evolves adiabatically and ends up in an eigenstate of the Hamiltonian. In the Two-Point Measurement (TPM) scheme~\cite{Kurchan2000,Tasaki2000}, the adiabatic evolution takes place between two projective measurements of the Hamiltonian at times $t_i$ and $t_f$ leading to measurement outcomes indicating the energies $E(t_i)$ and $E(t_f)$ so that the extracted work (positive when the system provides work) is defined by the change of energy $W=E(t_i)-E(t_f)$. Note that lifting the adiabatic assumption leads to an additional contribution in the change of energy coming from the exchange heat. This TPM scheme allows to recover thermodynamics fluctuation relations such as the Jarzynski equality in the case of classical information acquired on a quantum system.
\paragraph{}The two experiments of Masuyama \emph{et al.}~\cite{Masuyama2017} and Naghiloo \emph{et al.}~\cite{Naghiloo2018} both use a TPM scheme to infer the work exchanged between the qubit and the battery. Note that a strong assumption of this TPM scheme is the adiabatic nature of the evolution between projective measurements. For the above experiments, it requires that the operation time (about 0.01 to 1~$\mu$s) is much smaller than the thermalization time of the qubit, which is given by the qubit lifetime $T_1$ (about 10-100~$\mu$s).

\subsubsection{Weak and strong measurements}

\begin{figure}[b]
	\begin{center}
	\includegraphics[width = 0.4 \textwidth]{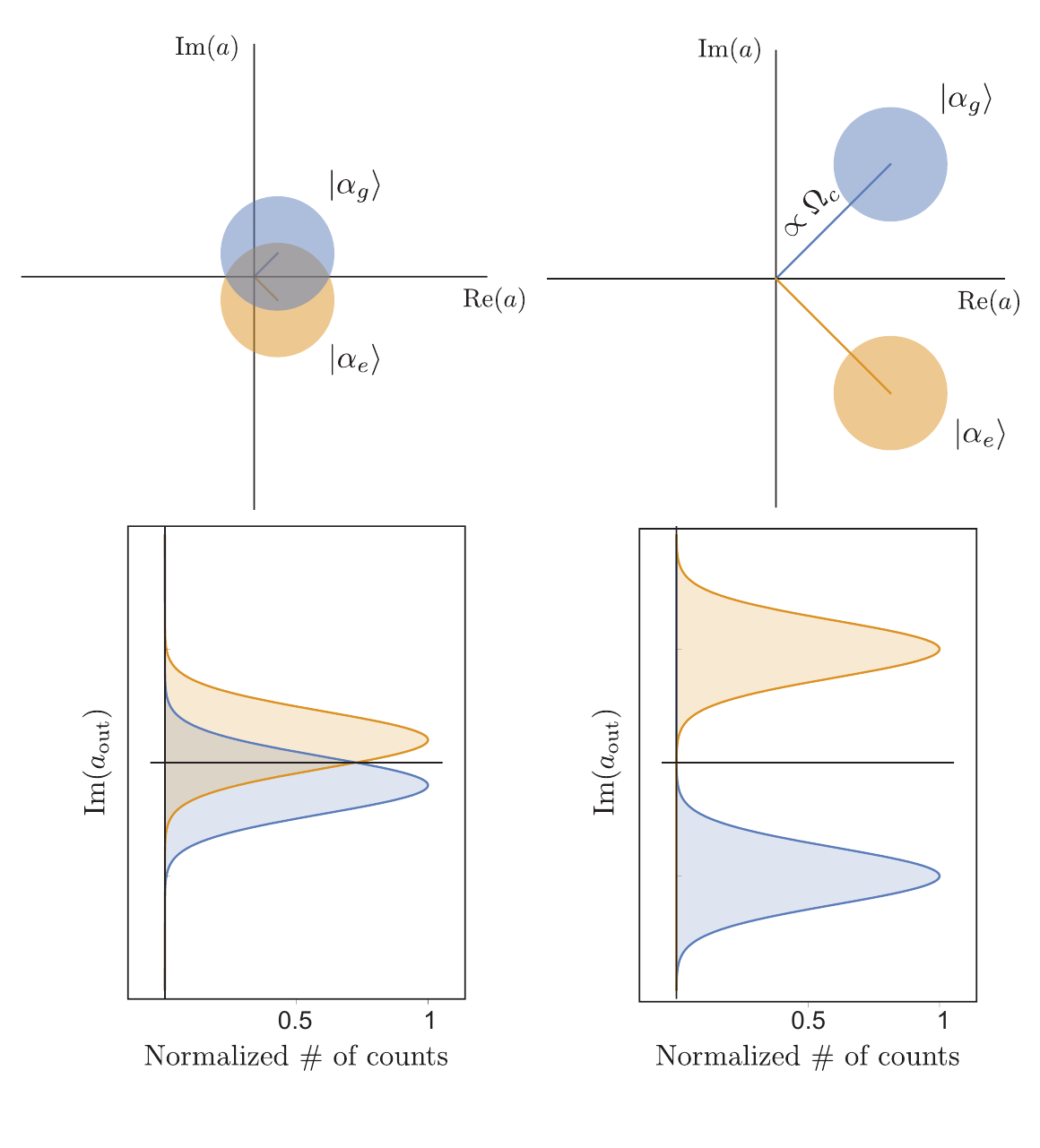}
\caption{\label{readout}Weak to strong measurement of a qubit. (top) When driven at resonance, the cavity hosts a coherent state $|\alpha_{g,e\rangle}$ that depends on the qubit state. Each disk represents the Gaussian distribution of the Wigner function of states $|\alpha_{g,e\rangle}$ in the phase space of the cavity mode quadratures. The disk radius corresponds to the vacuum fluctuations. (bottom) Histograms of the measurement outcomes for a detector that is sensitive to the quadrature encoding the qubit state in the cavity output field. When the drive amplitude $\Omega_c$ is small enough (left) the two states strongly overlap and the histograms are not well separated leading to a weak measurement. At larger $\Omega_c$ (right) the states and histograms are well separated and the measurement is strong.}
\end{center}
\end{figure}

We have discussed in the introduction on circuit-QED the way dispersive measurement operates. A big asset of circuit-QED for implementing a Maxwell demon is the possibility to arbitrarily tune the amount of information that the demon extracts from the qubit. This skill arises from the fact that the two cavity coherent states $|\alpha_{g,e}\rangle$ corresponding to the qubit in $|g\rangle$ or $|e\rangle$ are not orthogonal. More precisely, their overlap is $|\langle\alpha_e|\alpha_g\rangle|=e^{-|\alpha_e-\alpha_g|^2/2}$. When the cavity is coupled to a transmission line at a rate $\kappa$, the measurement rate~\cite{Clerk2008} , i.e. the rate at which information about the qubit state leaks towards the transmission line, is given by $\Gamma_m=\kappa|\alpha_e-\alpha_g|^2/2\propto\Omega_c^2$ where $\Omega_c$ is a drive strength that appears in equation~(\ref{Hc}) and is proportional to the drive amplitude. The measurement rate does not necessarily quantifies how much information is effectively acquired by the measurement apparatus at the other end of the transmission line. It just sets an upper bound by describing the case of a perfect measurement, and thus corresponds to the measurement induced dephasing rate. Taking into account the finite efficiency $0\leq\eta\leq1$ with which information is transmitted between the transmission line input and the final detector, one can more generally write the effective rate at which information about the qubit is acquired by the detector $\Gamma_m^\mathrm{eff}=\eta\Gamma_m$. For a given measurement duration $t_m$, the measurement is said to be weak (respectively projective) when $\Gamma_mt_m<1$ (resp. $\Gamma_mt_m\gg 1$). The strength of the measurement can be experimentally tuned by choosing the drive strength $\Omega_c$ (see Fig.~\ref{readout}).

\subsection{Probing quantum fluctuation theorems with weak measurements}
\subsubsection{Jarzynski for a discrete weak quantum measurement}
One of the main interest of the experiment of Masuyama \emph{et al.} is that it puts to the test~\cite{Masuyama2017} the Sagawa-Ueda quantum generalization of the Jarzynski equality where a so-called Quantum-to-Classical mutual information plays a key role (see Ref.~\cite{Funo2013} and \chqfluct). Varying the measurement strength at which the demon extracts information about the system allows them to tune this mutual information and provides a relevant test of the equality. The demon first performs a weak or strong measurement that leads to a measurement outcome $k$. A projective measurement is then performed, leading to some outcome $y$, right afterwards so that the system gets either to the ground or to the excited state. Based on the outcome $k$ alone, the demon then sends a feedback pulse to the qubit in order to try to extract a quantum of work out of it. Following the work of Funo, Watanabe and Ueda (see Ref.~\cite{Funo2013} and chapter~\chqfluct)  the quantity of information acquired by the demon during the measurement of outcome $k$ is given by the stochastic Quantum-to-Classical mutual information
\begin{equation}
\label{Iqc}
I_\mathrm{QC}(i,k,y)=\ln p(y|k)-\ln p(i),
\end{equation}
where $p(i)$ is the probability to get the outcome $i$ during the first projective measurement of the TPM that surrounds the whole pulse sequence, $p(y|k)$ is the probability to measure the outcome $y$ during the projective measurement conditioned on $k$. $I_\mathrm{Sh}(i)=-\ln p(i)$ is the stochastic Shannon entropy of the initially thermalized qubit. In the limit where the first measurement is strong and in the absence of decay of the qubit the two outcomes $k$ and $y$ match, therefore $p(y|k)=\delta_{y,k}$ and the stochastic mutual information $I_\mathrm{QC}(i,k,y)$ is simply given by the stochastic Shannon entropy corresponding to the first measurement of the TPM.
\paragraph{}In presence of feedback and when the initial and final Hamiltonian are identical, the work $W$ and the information $I_\mathrm{QC}$ extracted from the system by the demon verify the following generalized Jarzynski equality (see chapter~\chqfluct)
\begin{equation}
\label{jarz}
\langle e^{\beta W - I_\mathrm{QC}}\rangle=1-\lambda_\mathrm{fb}
\end{equation}
with $\beta=1/k_BT$ the inverse temperature. This equality takes into account the absolute irreversibility induced by the measurement operation of the demon. It is done via the probability $\lambda_\mathrm{fb}$ of irreversible events owing to the measurement. In the case of weak measurements irreversible events disappear because any forward events become possible, as unlikely as they can be. As a result $\lambda_\mathrm{fb}=0$ for weak measurements. The usual Jarzynski equality $\langle e^{\beta W}\rangle=1$ can thus be simply generalized to the case of non zero stochastic mutual information by replacing $W$ by $W-k_BTI_\mathrm{QC}$ in the equality.
\begin{figure}[H]
	\begin{center}
	\includegraphics[width = 0.4 \textwidth]{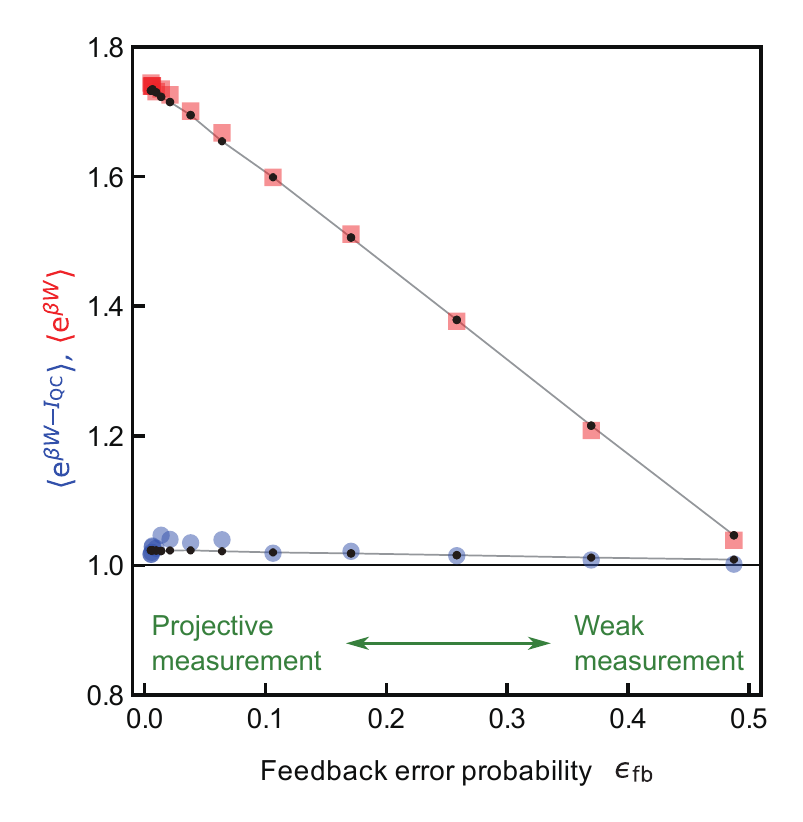}
\caption{\label{weak_jarz}(Adapted from Ref.~\cite{Masuyama2017}) Jarzynski equality is verified for any measurement strength only when the mutual information $I_\mathrm{QC}$ is taken into account. Here the blue dots correspond to the measured $\langle e^{\beta W - I_\mathrm{QC}}\rangle$ as a function of feedback error probability while red squares correspond to $\langle e^{\beta W}\rangle$.}
\end{center}
\end{figure}
\paragraph{}Fig.~\ref{weak_jarz} shows experimental data of $\langle e^{\beta W - I_\mathrm{QC}}\rangle$ and $\langle e^{\beta W}\rangle$ as a function of the feedback error probability $\epsilon_\mathrm{fb}$ for a qubit initialized with approximatively $10\%$ of excited state (from~\cite{Masuyama2017}). It is defined as the probability that the projective measurement outcome $y$ does not match the weak measurement outcome $k$: $\epsilon_\mathrm{fb}=p(y=e|k=g)+p(y=g|k=e)$. For strong measurement, the feedback process is almost error free ($\epsilon_\mathrm{fb}\ll 1$), while when the measurement gets weaker and weaker, the error goes towards $\epsilon_\mathrm{fb}=50~\%$. The latter value corresponds to the limit where the demon acts completely erratically because of its lack of information. The experiment shows that $\langle e^{\beta W - I_\mathrm{QC}}\rangle$ (blue dots) is almost equal to 1 no matter the strength of the measurement, while the uncorrected Jarzynski expression $ \langle e^{\beta W}\rangle$ (red squares) only reaches 1 when the feedback reaches its highest error probability. This effect can be simply understood by the fact that a $50~\%$ error probability means that no information is acquired by the demon and therefore $I_\mathrm{QC}=0$. In contrast, the situation when the measurement is so strong that it can be considered as projective might look surprising. According to equation~\ref{jarz}, when the feedback measurement is projective, one should expect irreversible events to appear, yielding $\lambda_\mathrm{fb}>0$ and implying $\langle e^{\beta W - I_\mathrm{QC}}\rangle<1$. Yet experimental data suggest otherwise, showing an average above one. The reason is to be found in qubit decay. First, as highlighted previously, the work has been assimilated to the energy change of the qubit in the TPM, resulting in a small overestimation of the work extracted from the qubit when it decays. Second, the qubit decay between the first TPM measurement and the two measurements $k$ and $y$ does not restrict forward processes and $\lambda_\mathrm{fb}$ stays null even in the strong measurement limit. Equivalently the qubit decay means that $p(y|k)$ is never strictly equal to $\delta_{y,k}$, and hence $I_\mathrm{QC}$ does not equal $I_\mathrm{Sh}$ even in the strong measurement limit.

\subsubsection{Jarzynski for continuous quantum measurements}
\begin{figure}[H]
	\begin{center}
	\includegraphics[width = 0.4 \textwidth]{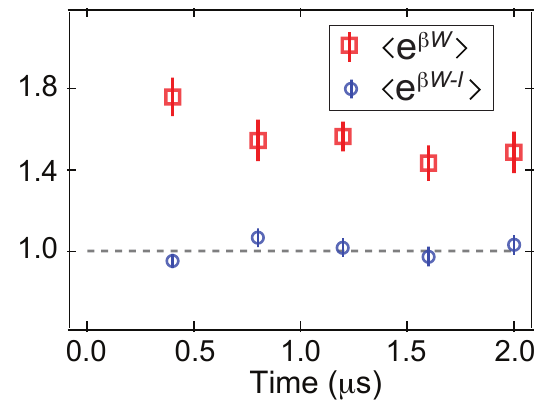}
\caption{\label{traj_jarz}(Adapted from Ref.~\cite{Naghiloo2018})Jarzynski equality is verified for any measurement time only when the mutual information $I_\mathrm{QC}$ is taken into account. Here the blue dots correspond to the experimentally inferred $\langle e^{\beta W - I_\mathrm{QC}}\rangle$ as a function of measurement time $t_m$ while red squares correspond to $\langle e^{\beta W}\rangle$.}
\end{center}
\end{figure}

In the previous part, the weak measurement provides a single measurement outcome $k$ on which the feedback is conditioned. In all generality, the measurement record can be a continuous signal $\{V(t)\}_{0<t<t_m}$ that lasts for some total duration $t_m$. Then how can the demon optimally extract work from the system and how to quantify the knowledge of the demon about the system? This is what the experiment of Naghiloo \emph{et al.}~\cite{Naghiloo2018} addresses. It is in fact possible to infer the qubit state $\rho_t$ at any time conditioned on the continuous measurement record (see Appendix and Ref.~\cite{campagneibarcq:tel-01248789}). This is called a quantum trajectory. Importantly, the conditional density matrix $\rho_{t_m}$ at the end of the measurement encodes everything one needs to know to predict the statistics of any following measurement on the qubit. In their experiment, Naghiloo \emph{et al.} chose to drive the qubit during the measurement so that $\sigma_X$ is non zero during the quantum trajectory. The information acquired by the demon then takes into account the fact that the demon not only has knowledge on the qubit energy expectation in $\sigma_Z$ as in the previous experiment but also in the qubit coherence in $\sigma_X$. The density matrix can always be written as $\rho_{t_m}=p_1|\psi_{t_m}\rangle\langle\psi_{t_m}|+p_0|\psi_{t_m}^\perp\rangle\langle\psi_{t_m}^\perp|$ for one particular pure qubit state $|\psi_{t_m}\rangle$ and its orthogonal one $|\psi_{t_m}^\perp\rangle$. Note that due to the limited efficiency of the detector (here $\eta=30~\%$), the quantum states are mixed and $p_1,p_0<1$. 
In order to optimally extract work out of the qubit, the demon needs to perform a pulse at the qubit frequency that brings $\rho_{t_m}$ to $\max(p_0,p_1)|g\rangle\langle g|+\min(p_0,p_1)|e\rangle\langle e|$. In their experiment, Naghiloo \emph{et al.}~\cite{Naghiloo2018} avoid the complexity of calculating the proper pulse to send in real time by performing rotations around the $y$ axis of the Bloch sphere with a random angle and then postselecting the right ones by postprocessing.
\paragraph{} As we have seen above, in the case of a quantum system and a classical demon such as here, the fluctuation relation needs to take into account the stochastic mutual information $I_{QC}$. This quantity is determined in a slightly different manner from for a discrete weak measurement. If one were to perform an ideal projective measurement at time $t_m$ of the observable $|\psi_{t_m}\rangle\langle\psi_{t_m}|$, one would get an outcome $z'=1$ with probability $p_{t_m}(z'=1)=p_1$ and $z'=0$ with probability $p_{t_m}(z'=0)=p_0$. With similar notations, a projective measurement of the observable $|e\rangle\langle e|$ after the qubit is thermalized at the beginning of the experiment (time $0$) leads to an outcome $z=1$ with probability $p_0(z=1)=(1+e^{\beta\hbar\omega_q})^{-1}$ and $z=0$ with probability $p_0(z=0)=(1+e^{-\beta\hbar\omega_q})^{-1}$. The stochastic mutual information~\cite{Funo2013} can then be written as
\begin{equation}
I_{QC}(z,z')=\ln p_{t_m}(z')-\ln p_0(z).
\label{Itraj}
\end{equation}
The above expression is very similar to Eq.~(\ref{Iqc}). This illustrates that the main difference between the experiments of Masuyama \emph{et al.} and Naghiloo \emph{et al.} is not so much in the discrete versus continuous measurement approach since in the end only the last quantum state $\rho_{t_m}$ matters. It is in the fact that the first focuses on states that do not have any quantum coherence while the second extends the experiment to finite coherences by adding a drive during the measurement.
\paragraph{}Concretely, after initialization in a thermal state, Naghiloo \emph{et al.} perform the first TPM projective measurement and measure continuously and dispersively the qubit state while a drive resonating at the qubit frequency induces a rotation of the qubit in the Bloch sphere. A feedback pulse is then applied by a postselection based on the reconstructed quantum trajectory. The sequence is terminated by the second TPM measurement. While the strength of the measurement allowed to tune the stochastic mutual information in the work of Masuyama \emph{et al.}, here the authors decided to keep a constant measurement rate and vary the duration $t_m$ of the measurement. Noting that the free energy difference between the initial and final states is zero because the Hamiltonian is the same, the extracted work $W$ and demon information verify:
\begin{equation}
\label{jarz_traj_formulat}
\begin{split}
\sum_{z,z'\in\{0;1\}}p_0(z)p_{t_m}(z')e^{\beta\hbar\omega_q(z-z')-I_{QC}(z,z')}\\
=\langle e^{\beta W-I_{QC}}\rangle=1\ .
\end{split}
\end{equation}
Experimentally, the work $W$ that is extracted both during the measurement under a drive and during the feedback pulse is determined using the TPM protocol. The  inferred evolution of $\langle e^{\beta W-I}\rangle$ and $\langle e^{\beta W}\rangle$ are represented in Fig.~\ref{traj_jarz} as a function of the measurement duration $t_m$ for a qubit initially at equilibrium at a temperature $\hbar\omega_q/4k_B$. As in the experiment by Masuayama \emph{et al.}, the generalized Jarzynski equality is indeed verified. This demonstrates that the feedback pulse is indeed applied the right way and validates the definition of information.

\subsection{Information loss during weak measurements}
Beyond Jarzynski equalities, the information acquired by the demon exposes the differences between a weak measurement in the $Z$ direction with or without a drive at the qubit frequency.
\paragraph{}Masuyama \emph{et al.} show the evolution of $\langle I_\mathrm{QC}\rangle$ as a function of the feedback error probability $\epsilon_\mathrm{fb}$. 
When the collected information is maximal (or equivalently the feedback error probability is zero) the mutual information does not quite reach $\langle I_\mathrm{Sh}\rangle$ due to the finite decay of the qubit. As expected the average of the demon information decays to zero as the feedback error probability goes up. 
\paragraph{}Similarly Naghiloo \emph{et al.} compute the information acquired by the demon in the case of a continuous measurement when the qubit starts in a thermal state $\rho_0$. Since the qubit is not actually projectively measured after time $t_m$, one needs to sum over the different possible outcomes $z$ and $z'$ and gets for a single measurement record $\{V(t)\}_t$
\begin{equation}
\label{info_traj}
\begin{split}
\langle I_{QC}\rangle_{\{V(t)\}_t}&=\sum_{z,z'\in\{0;1\}}p_0(z)p_{t_m}(z')I_{QC}(z,z')\\&=S(\rho_0)-S(\rho_{t_m}).
\end{split}
\end{equation}
where $S(\rho)=-\mathrm{Tr}(\rho\ln\rho)$ denotes the Von Neumann entropy. The information acquired by the demon over a trajectory is hence simply the difference between the initial and final entropies of the qubit. When the quantum efficiency $\eta$ is 1, no information about the system is lost during the continuous measurement. The qubit state that is reconstructed from the measurement records using the stochastic master equation~(\ref{SME}) conserves its initial purity. Since experimentally $\eta\approx30\%$, information is lost in the environment. If the lost information is larger than the gained information during the measurement one gets $\langle I_{QC}\rangle_{\{V(t)\}_t}<0$. This is the case when the initial state is close to a pure state: during the trajectory the state loses its coherence and hence purity because of the imperfection of the measurement, increasing its entropy. On the other hand when the initial state is close to the most entropic state, the measurement purifies the state of the qubit and the final entropy becomes smaller than the initial one. This transition has been experimentally observed and can be seen in Ref.~\cite{Naghiloo2018} where the mutual information goes from positive to negative values.

\section{AUTONOMOUS DEMON}
\subsection{Coherent information transfer and work extraction}
In the two experiments presented previously the information has to leave the quantum world to be recorded and used in the feedback process of the demon. Yet, it is possible to realize a fully quantum experiment where the demon itself is a quantum system. This is a case where the control is deterministic and unconditional hence without any feedback based on measurement. Instead built-in conditional operations need to be designed for the demon to operate by autonomous feedback. In the dispersive Hamiltonian~(\ref{H}), when $\chi$ is larger than the linewidths of both the qubit and cavity, a regime known as \emph{photon-resolved}~\cite{Schuster2007}, it becomes possible for a pulse at a given frequency to excite the cavity (respectively qubit) conditionally on the number of excitations in the qubit (respectively cavity). More precisely a drive at $\omega_c$ displaces the cavity only if the qubit is in its ground state, while a drive at $\omega_q-N\chi$ flips the qubit only when the cavity hosts exactly $N$ photons.
\paragraph{}This autonomous quantum Maxwell demon was realized in Ref.~\cite{Cottet2017}. Initially we assume the cavity to be in the vacuum state $|0\rangle$ as thermal excitations can be neglected. After initialization of the qubit in a thermal or in a superposition of energy eigenstates, a pulse at $\omega_c$ is applied with a duration chosen to be larger than $\chi^{-1}$ to ensure selectivity. The cavity thus ends up either in $|\alpha_e\rangle=|0\rangle$ if the qubit is excited, or ideally in a coherent state $|\alpha_g\rangle=|\alpha\rangle$ if the qubit is in the ground state. Since the process follows a unitary evolution, an initial superposed state like $(|e\rangle+|g\rangle)/\sqrt{2}$ results in an entangled state $(|e\rangle|0\rangle+|g\rangle|\alpha\rangle)/\sqrt{2}$. Consecutively, a $\pi$-pulse at $\omega_q$ flips the qubit only if the cavity hosts 0 photon. It is always the case when the qubit is excited, and it happens with a probability $|\langle0|\alpha\rangle|^2=\exp(-|\alpha|^2)$ when the qubit is in the ground state. If $\alpha\gg 1$, the demon distinguishes well between ground and excited states and the qubit always ends up in the ground state. Consequently, the information about the qubit state makes the energy exchange of 1 quantum of work between the drive pulse and the qubit unidirectional: the drive is either reflected without loss of energy (if the qubit has no energy to offer) or contains one extra stimulated-emitted photon (if the qubit is in $|e\rangle$). A positive net extraction of work is thus ensured at $\alpha\gg 1$. In the case of an initially superposed qubit, the conditional $\pi$-pulse disentangles the qubit and the cavity so that the cavity ends up a state $(|0\rangle+|\alpha\rangle)/\sqrt{\mathcal{N}}$. Therefore the conjugation of conditional displacement and $\pi$-pulse swaps the qubit and cavity states and performs a coherent information transfer from the qubit to the cavity. On the other hand if $\alpha$ is not large enough, the conditional $\pi$-pulse does not fully disentangle the qubit and the cavity, the information transfer is imperfect and as a result the work extracted is smaller. Similarly to what has been done in Ref.~\cite{Masuyama2017}, the quantity of information transferred to the demon can be tuned by varying the amplitude of the displacement $\Omega_c$ or, equivalently, the mean number of photons in the cavity $\bar{n}$. 
\subsection{Information transfer}
\begin{figure}[H]
	\begin{center}
	\includegraphics[width = 0.4 \textwidth]{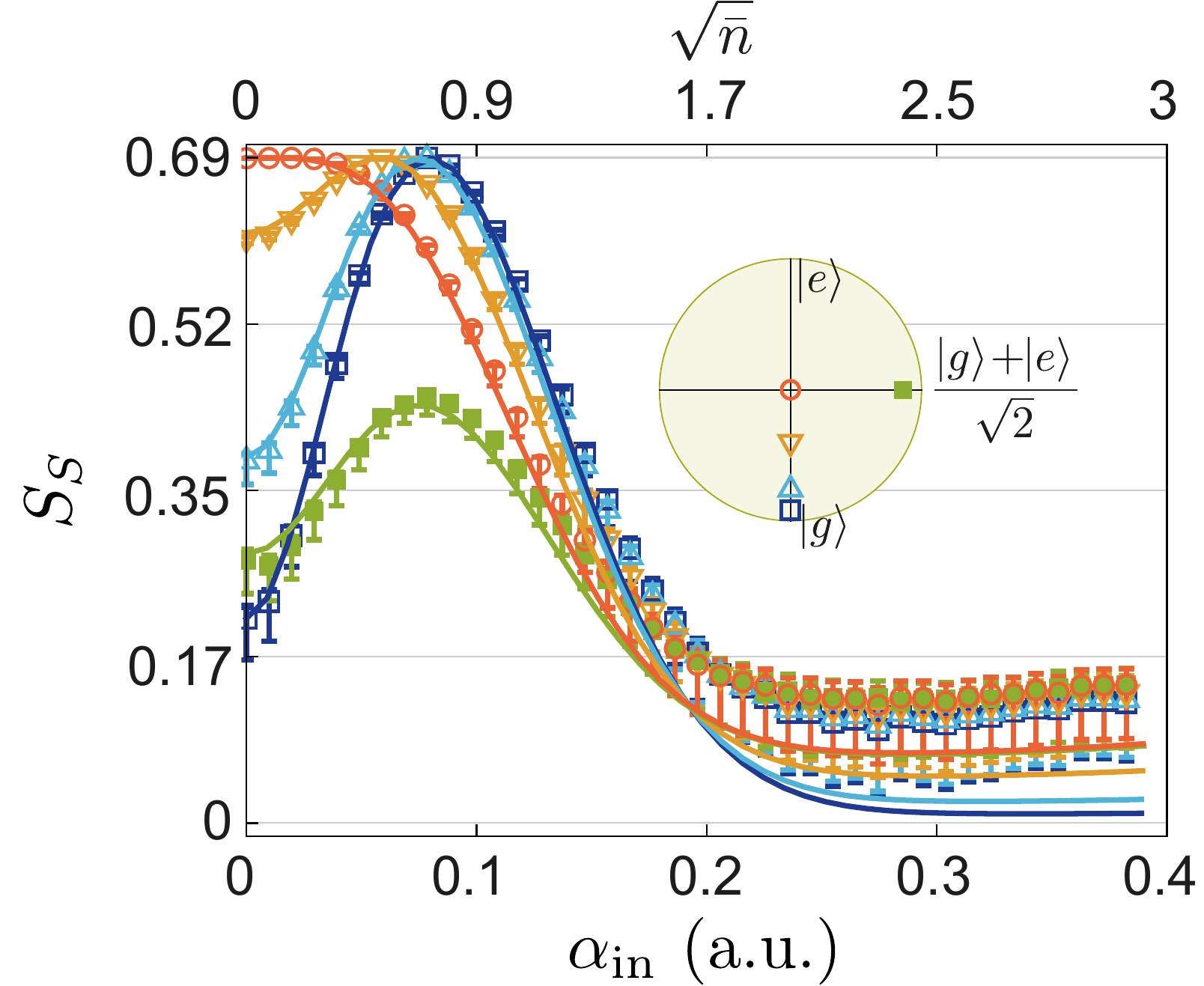}
\caption{\label{S_q}(Figure from Ref.~\cite{Cottet2017}) Final measured Von Neumann entropy of the system as a function of the amplitude of the cavity displacement drive amplitude (labeled $\alpha_\mathrm{in}$), which tunes the information quantity that the demon extracts from the qubit.}
\end{center}
\end{figure}

\begin{figure}[H]
	\begin{center}
	\includegraphics[width = 0.4 \textwidth]{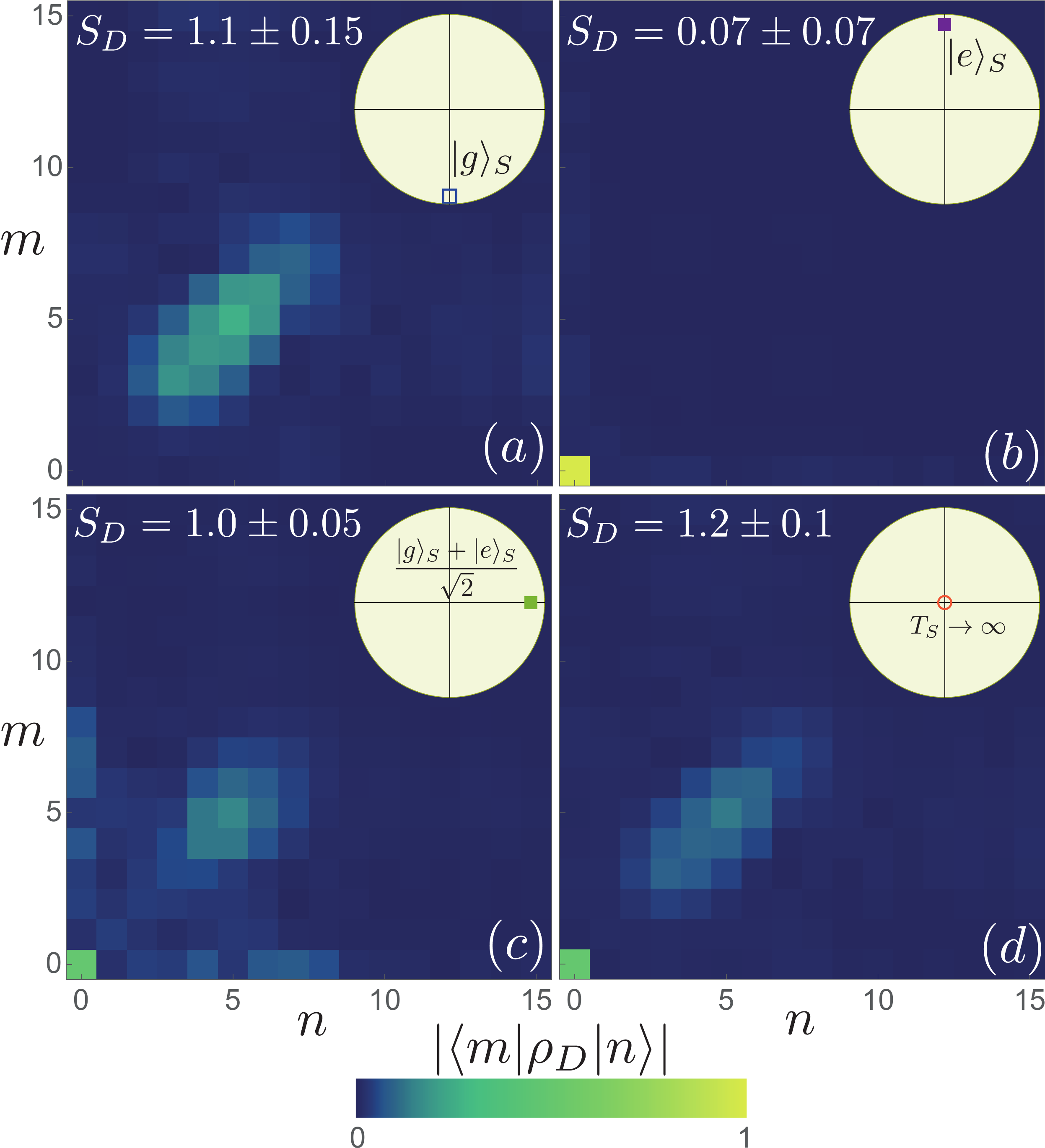}
\caption{\label{S_d} (Figure from Ref.~\cite{Cottet2017}) Amplitude of the elements of the reconstructed density matrix of the cavity (demon's memory) in the Fock states basis after the sequence for a qubit initialized in the (a) ground state, (b) excited state, (c) superposed state and (d) thermal state at infinite temperature.}
\end{center}
\end{figure}

In our work~\cite{Cottet2017}, the whole sequence is terminated by a full tomography of the final qubit (system) state $\rho_S$ using a set of projective measurements. The evolution of the final Von Neumann entropy of the qubit $S_S=-\mathrm{Tr}(\rho_S\ln \rho_S)$ with $\sqrt{\bar{n}}$ is represented in Fig.~\ref{S_q} for various initial states of the qubit, either in a thermal or superposed state. Its evolution exhibits a clear quantum feature that highlights the quantumness of the demon. The entropy of the qubit first goes to a maximum before eventually decreasing.  This increase of entropy manifests the residual presence of entanglement between the qubit and the cavity after the work extraction pulse: when measuring the state of the qubit only, one discards the information encoded in the cavity and gets a more entropic qubit. Within the interpretation of the experiment in terms of Maxwell's demon, this large qubit entropy means that the demon operates erratically due to the partial quantum information it gets on the qubit. It is not the case when $\bar{n}\ll 1$, because then the conditional $\pi$-pulse at $\omega_q$ is always on resonance. The behavior of the demon becomes perfectly predictable and does not affect the entropy of the qubit. In the limit of large $\bar{n}$ however, the information transfer is large enough so that the demon lowers the entropy of the qubit. The residual entropy is mostly due to the parasitic thermalization of the qubit with the environment during the sequence.
\paragraph{}When the demon's memory is a quantum system, such as the cavity here, it becomes possible to realize a full quantum tomography of its state and to uncover its quantum coherences. In our experiment~\cite{Cottet2017}, we used the qubit to perform a tomography of the cavity at the end of the sequence based on generalized Husimi Q-functions measurement and state reconstruction~\cite{Kirchmair2013a}. Because we used a single qubit as the system and as a tomographic tool, it is necessary to be certain that the qubit is in the ground state before starting the tomography. For that reason, the range of cavity displacement amplitudes where one can reconstruct the state of the cavity is limited to the cases where the demon cools efficiently the qubit close to its ground state. In particular, the technique does not allow us to measure the variations of the demon's information with the displacement drive amplitude. Such a measurement would be possible if one would use another ancillary qubit just to perform the cavity tomography~(for instance with an architecture as in Ref.~\cite{Blumoff2016}). The magnitude of the elements of the reconstructed density matrix of the cavity $\rho_D$ is represented in Fig.~\ref{S_d} in the Fock state basis for 4 different initial qubit states: (a) ground, (b) excited, (c) superposed and (d) thermal at infinite temperature. As expected, the cavity contains a large number of photons when the qubit is initially in the ground state ($\bar{n}\approx4.6$) and stays in vacuum when the qubit is initialized in the excited state. The coherence of the process arises when comparing the superposed case and the thermal one. When the qubit is initially superposed, the cavity ends up with non-zero off-diagonal terms of the form $\langle 0|\rho_D|m\rangle$, $m\in \mathbb{N}^*$, showing coherences between the vacuum and the displaced state. These off-diagonal terms are zero in the thermal case.

\paragraph{}From the reconstructed density matrix of the demon $\rho_D$ one can infer its Von Neumann entropy in an attempt to quantify the amount of information stored in the cavity. It is indicated on Fig.~\ref{S_d}. Surprisingly its state is very entropic except when it is in vacuum. This is due to the conjugated effect of the unwanted qubit induced non-linearity in the cavity and of dissipation. As a result the displacement produces an entropic state instead of a coherent state. Yet, the comparison of the superposed and thermal cases shows that the cavity entropy in this case indeed reflects the initial entropy of the qubit, with the superposed state resulting in a less-entropic cavity than the thermal one. However, besides highlighting quantum effects in the transfer of information, it is not possible to perform an information balance between the qubit and the cavity. This is only due to the fact that the information on the two-dimensional qubit is encoded in a multi-level system, the cavity, which does not remains pure due to the aforementioned parasitic nonlinearities. With this type of encoding, the only relevant information for the demon to operate efficiently is whether or not the cavity is in $|0\rangle$. Therefore, one could think of a better definition for the stored information in such a way that it would eliminate the irrelevant contribution of entropy in the excited states of the cavity.

\subsection{Direct work measurement}
Interestingly, it is possible to perform a direct measurement of the work extracted by the demon without resorting to the TPM process. It is done by directly recording the power contained in the reflected $\pi$-pulse. Given the small energy of a single microwave photon, the use of quantum limited Jopsephson amplifiers was instrumental~\cite{Roy2018}. To do so in Ref.~\cite{Cottet2017}, we amplify the reflected pulse by a Josephson Parametric Converter (JPC)~\cite{Roch2012}. It amplifies the two field quadratures by the same amount and as a result acts as a \emph{phase-preserving} amplifier. After amplification the field is digitized and the average instantaneous power at $\omega_q$ is extracted. Denoting as $a_\mathrm{in,out}$ the annihilation operator of a photon propagating in the transmission line towards (respectively from) the cavity, one can simply express the power extracted from the qubit $P_\mathrm{ext}$ by the difference between the photon rate that is sent and the one that is reflected $P_\mathrm{ext}=\hbar\omega_q(\langle a_\mathrm{out}^\dagger a_\mathrm{out}\rangle - \langle a_\mathrm{in}^\dagger a_\mathrm{in}\rangle)$. Besides one can write the propagating number of photons in the transmission line in terms of qubit operators~\cite{Cohen-Tannoudji2001}
\begin{equation}
\label{Nphotons}
\frac{P_\mathrm{ext}}{\hbar\omega_q}=\gamma_a\frac{1+\langle\sigma_z\rangle}{2}+\frac{\Omega_q}{2}\langle\sigma_x\rangle,
\end{equation}
\begin{figure}[H]
	\begin{center}
	\includegraphics[width = 0.9 \textwidth]{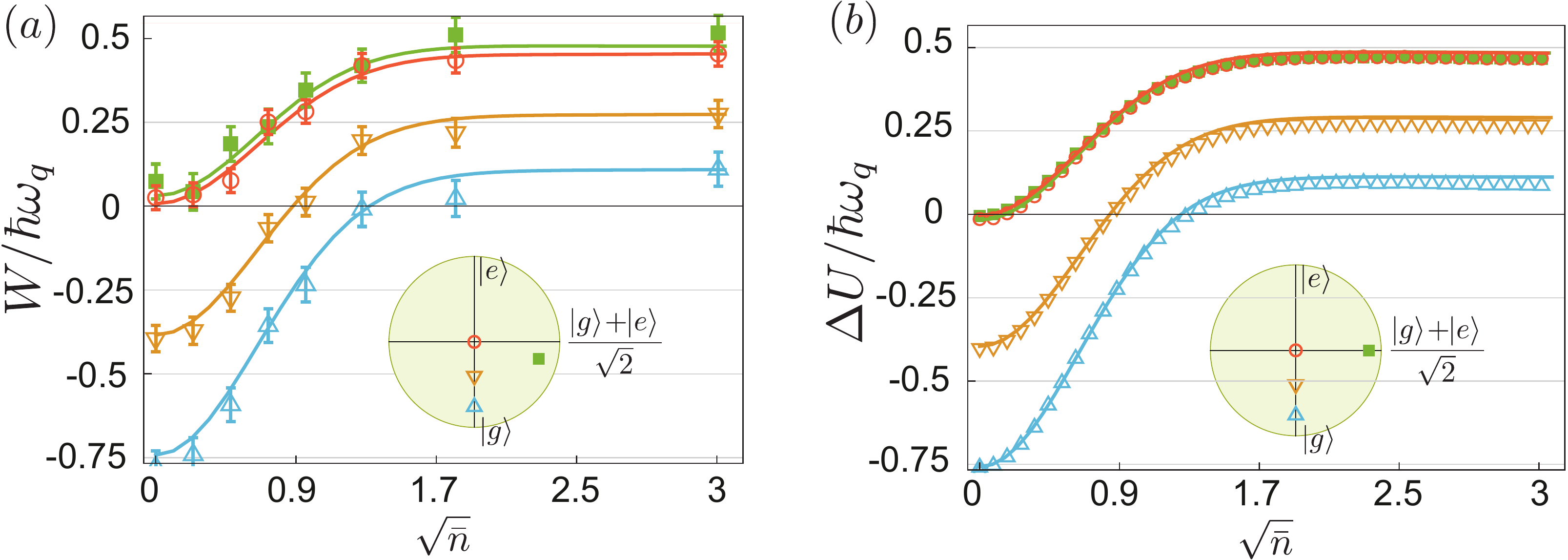}
\caption{\label{work} (Adapted from~\cite{Cottet2017})(a) Directly measured extracted work from the qubit by the autonomous demon for various initial thermal or superposed states. The direct measurement perfectly agrees with the (b) independent measurement of the variation of energy of the qubit between two projective measurements (TPM).}
\end{center}
\end{figure}
where $\gamma_a$ is the emission rate of the qubit towards the transmission line and $\Omega_q$ is the Rabi frequency as defined in equation~\ref{Hq}. The first term is proportional to the probability to find the qubit in the excited state and thus corresponds to spontaneous emission of the qubit. To understand the second one, let us integrate it over half a Rabi oscillation. In the absence of losses ($\Omega_q\gg\gamma_a$), if the qubit goes from $|g\rangle$ to $|e\rangle$, one would find $-1$, and $+1$ for a qubit going from $|e\rangle$ to $|g\rangle$. Therefore, in case of negligible loss during the pulse, this term quantifies the coherent energy exchanges between the drive and the qubit through absorption/stimulated emission cycles.
\paragraph{}Experimentally the reflected signal is decomposed into short time bins and integrated to get the extracted work $W=\int_0^{\pi/\Omega_q}P_\mathrm{ext}\mathrm{d}t$. The evolution of the extracted work with the number of photons contained in the demon is represented in Fig.~\ref{work} as well as the variation of the qubit internal energy $\Delta U=E(t_i)-E(t_f)$ obtained from the qubit tomography, which is equivalent to a TPM in the case of projective measurement along $\sigma_Z$. As expected the extracted work is negative when the demon does not distinguish the ground from the excited state, and becomes positive as the number of photons in the cavity increases. The direct work measurement and the qubit energy change measurement are not obtained for the exact same pulse sequences due to technical issues in the work measurement (see supplementary material of~\cite{Cottet2017}). A consequence is that the superposed state prepared in the two cases is not exactly the same, as represented in the Bloch sphere projections in the $XZ$-plane in insets. Nevertheless the agreement between the two independent measurements is remarkable.

\section{CONCLUSION}
\subsection{Summary on already realized experiments}
The three experimental realizations of a quantum Maxwell's demon presented in this chapter demonstrate that superconducting circuits constitute a useful and versatile testbed for quantum thermodynamics. They make possible the experimental validation of thermodynamical equalities in the context of various measurements: strong, weak, quantum trajectories or coherent transfer to an ancillary quantum system. The main features of these three experiments are shown in table~\ref{summary}. On the basis of this table, we can foresee  directions towards which future experimental realizations with superconducting circuits could go. A direct work measurement (using direct microwave measurement of the released energy~\cite{Cottet2017}, calorimeters~\cite{Pekola2012} or other techniques) coupled to a demon using classical information would give access to the influence of irreversible events and how they arise when the measurement becomes strong. Coupled to quantum trajectory measurements, a direct work measurement through fluorescence would allow one to precisely quantify and separate the flows of heat and work during the trajectory and the feedback pulse. Finally in the case of quantum demon memories, the use of an ancillary qubit would allow one to perform joint measurements of the states of the qubit and cavity, and hence to quantify the mutual information between them at any measurement strength. This would lead to the experimental measurement of a fully quantum Jarzynski equality in the presence of quantum coherence.
\begin{table*}[h]
\centering
\begin{tabular}{|c|c|c|c|}
\hline
 & Masuyama \emph{et al.}~\cite{Masuyama2017} & Naghiloo \emph{et al.}~\cite{Naghiloo2018} & Cottet, Jezouin \emph{et al.}~\cite{Cottet2017} \\ \hline
Nature of the demon & Classical & Classical & Quantum \\ \hline
Mutual information knob & Cavity drive amplitude & Measurement duration & Cavity drive amplitude \\ \hline
Work determination & Inferred from TPM & Inferred from TPM & Direct power measurement \\ \hline
Fluctuation relation & Sagawa-Ueda equality for & Jarzynski equality with & No measure of\\
				   &weak to strong measurement			& quantum trajectories	 & mutual information \\ \hline
Information evolution & Classical information & Information loss & Differing demon entropies for\\
					&						&	causing $\langle I\rangle \leq 0$&thermal and quantum cases \\ \hline

\end{tabular}
\caption{\label{summary}Status of the three existing experiments on quantum Maxwell's demon using superconducting circuits at the time of this writing.}
\end{table*}
\subsection{Theoretical proposals}
Over the past ten years few theoretical proposals have designed experiments based on superconducting circuits to probe further the physics of Maxwell's demon in quantum mechanics. First, it has been suggested to use superconducting circuits to realize a quantum Otto engine~\cite{Quan2006,Quan2007}, by using the tunability of the coupling between two charge qubits. A second important use of superconducting circuits is the possibility to quickly tune their frequency, allowing  a demon to extract work from the system up to the Landauer bound $k_B T\ln 2$. Such as scheme was proposed by Pekola \emph{et al.}~\cite{Pekola2016} for a single qubit using Landau-Zener transitions and measurement based feedback. In the three existing experiments, the energy levels of the qubit were fixed and it was thus  impossible to saturate the bound. More precisely the maximal work extracted by the demon from a fixed-frequency qubit is equal to the initial mean energy of the qubit: $W_\mathrm{max}=\hbar\omega_q p_e$ where $p_e$ denotes the initial probability to find the qubit in the excited state. At thermal equilibrium at temperature $T$ it reads $p_e=1/(1+\exp(\hbar\omega_q/k_BT))$ and the ratio of work over Landauer bound reads at best
\begin{equation}
\label{landauer}
\frac{W_\mathrm{max}}{k_BT\ln2}=\frac{\hbar\omega_q/k_BT}{\ln2(1+e^{\hbar\omega_q/k_BT})}
\end{equation}
and reaches a maximum determined numerically around $40\%$. However allowing to tune the energy levels of the qubit during the sequence leads to saturating the bound. First, the initial sequence is left unchanged: the qubit is thermalized at fixed frequency $\omega_1$, then measured and flipped to the ground state (whether or not the demon is classical does not change the energy balance here). As already stated, this technique extracts at best a work $W_1=\hbar\omega_1p_e$. Second, the energy levels are shifted adiabatically to a frequency $\omega_2$ so that $\hbar\omega_2\gg k_BT$. Since the qubit is in the ground state, this process can be done without any expense of external work. Third, the energy levels are brought back quasi-statically to the initial frequency $\omega_1$. This has to be done slowly enough so that the qubit is always at equilibrium with the heat bath at temperature $T$. This process extracts a work $k_BT\ln2-W_1$ and hence the total work extracted from the qubit reaches the bound. More generally, using tunable qubits leads to designing thermal machines able to operate at the Carnot efficiency. Josephson junction circuits offer such a tunability by the application of an external magnetic flux through a loop of two junctions~\cite{Wendin2016,Gu2017}. Various theoretical proposals suggest to use this tunability to perform Otto thermal machines operating either as engines or refrigerators~\cite{Niskanen2007,Karimi2016}. Owing to the possibility to perform single-shot measurements, superconducting circuits could exhibit the role of information transfers in such systems.
\paragraph{}In most of proposed realizations of Maxwell's demon, the information about the system is used to extract work from a single heat bath. This process is not the only apparent violation of the second principle: an information-powered refrigerator would apparently violate the second principle as well. Campisi \emph{et al.}~\cite{Campisi2017} proposed to use superconducting circuits coupled to calorimetric measurements to generate an inverse heat flow from a cold bath to a hot one. Each bath is modeled as an RLC resonator with a frequency that can be tuned by an external flux and at temperatures $T_{h,c}$, with $T_c<T_h$. Both resonators are inductively coupled to the same superconducting qubit. Recently developed calorimeters~\cite{Pekola2010,Govenius2014,Gasparinetti2015} allow to detect when an excitation leaves or enters each resistor, and provide the information acquired by the demon. Initially the cold bath and the qubit in its ground state are on resonance and the hot bath is far from resonance, inhibiting the effective coupling between the bath and the qubit. A calorimeter can detect the transfer of an excitation from the cold bath to the qubit. Such an event triggers, by measurement based feedback, pulses on two control fluxes that bring the hot bath in resonance with the qubit and put the cold bath out of resonance. The qubit eventually releases its excitation into the hot bath. This event is detected by the heating of the hot resistor through a second calorimeter and a second feedback control is applied to bring the system back to its initial state, hence closing the thermodynamic cycle.
\paragraph{}When measuring the state of the qubit, it is interesting to consider cases where the demon does not measure its state in the energy basis, i.e. along the $z$-axis of the Bloch sphere. A measurement that would project the qubit in a coherent superposition~\cite{Hacohen-Gourgy2016,Vool2016} would instead allow for more work extraction than classically allowed, by using the quantum coherence as a resource. Elouard \emph{et al.}~\cite{Elouard2017b} studied the case of a demon measuring the state of a superconducting circuit in the $x$-direction. The qubit is initialized in $(|g\rangle+|e\rangle)/\sqrt{2}$ then measured strongly along the $x$-axis in a stroboscopic way, and as a result is projected each time onto $(|g\rangle\pm|e\rangle)/\sqrt{2}$. Between the measurements the qubit is driven on resonance during a time $\tau$ with a Rabi frequency $\Omega_q$ and extracts positive work from the qubit if it was initially in $(|g\rangle+|e\rangle)/\sqrt{2}$. In the limit where $\Omega_q \tau\ll1$, the qubit has almost not evolved between two consecutive measurements and is re-projected with a very high probability on $(|g\rangle+|e\rangle)/\sqrt{2}$ by Zeno effect. Therefore the external pulse is continuously powered-up by the projective measurement of the qubit along $x$. Importantly, such a heat engine can be done in the absence of an external cold bath: the energy is here provided by the back-action of the measurement apparatus.

\subsection{Perspectives}
  The first experimental realizations of quantum Maxwell's demons in this platform illustrate the many possibilities offered by superconducting circuits. They pave the way to various more experiments that will explore the intimate link between information and thermodynamics in the quantum regime. With this goal in mind, one could think of using other kinds of systems than transmon qubits as  working agents. Among them, fluxonium qubits~\cite{Manucharyan2009} appear as an extremely promising platform because they offer a whole zoology of transitions. Their transition frequency can be tuned from hundreds of MHz to about $20$ GHz using an external magnetic flux, offering the possibility to study regimes where the system dynamically goes from $\hbar\omega\gg k_B T$ to $\hbar\omega\ll k_B T$. Moreover the coupling rates of fluxonium qubits with their environment can vary over 5 orders of magnitude, allowing one to finely engineer the heat exchanges with the baths. Superconducting circuits can also provide components of more sophisticated experiments that would use heat switches~\cite{Ronzani2018}. In a broader picture, the use of hybrid systems formed by superconducting circuits coupled to mechanical resonators appears as extremely promising. It would allow one to proceed to a work extraction that would indeed be used to lift a small mass, as in the first early descriptions of Maxwell's gedanken experiment. This could be interesting to solve controversies about the nature of heat and work in quantum systems. Superconducting circuits are also a promising platform for realizing entanglement between two qubits by the use of thermal baths only and in the absence of any coherent drive~\cite{BohrBrask2015}. With the steady improvement of superconducting qubits, there is no doubt that these systems offer a growing number of possibilities to test quantum thermodynamic properties and implement potential applications.
\bigskip

\acknowledgements
\section{ACKNOWLEDGMENTS}
This work was funded by Agence Nationale de la Recherche under the grant ANR-17-ERC2-0001-01. Part of the research reviewed in this chapter was made possible by the COST MP1209 network “Thermodynamics in the quantum regime".

\appendix
\section{APPENDIX}
\subsection{Quantum trajectories}
We describe here how one can determine the quantum state $\rho_t$ in the experiment of Naghiloo \emph{et al.}~\cite{Naghiloo2018}. The signal $V(t)$ carrying the information being continuous, one can decompose the information into time bins and get a time-resolved measurement. In order to observe non negligible effects of measurement backaction, it is crucial to minimize the amount of information that is lost between the cavity and the measurement apparatus, i.e. to maximize the quantum efficiency $\eta$. This process is done using quantum-limited amplification right after the cavity, a process that adds the minimal amount of noise allowed by quantum mechanics~\cite{Caves2012a}. In their quantum trajectories experiment, Naghiloo \emph{et al.} use a Josephson Parametric Amplifier (JPA)~\cite{Castellanos-Beltran2007} that amplifies one of the two quadratures of the field (equivalent to the position and momentum operators of a mechanical linear oscillator) and as such operates as a \emph{phase-sensitive} amplifier allowing to reach a state of the art quantum efficiency of about $\eta=30\%$. Denoting as $\mathrm{d}t$ the time interval between two successive records, the measurement record at time $t$ is given by $\mathrm{d}V(t)=\sqrt{2\eta\Gamma_m}\langle\sigma_z\rangle_{\rho_t}\mathrm{d}t+\mathrm{d}\mathcal{W}_t$ where $\mathrm{d}\mathcal{W}_t$ a zero-mean Wiener process whose variance is $\mathrm{d}t$ and which represents the quantum and technical noise. Note that here the expectation value $\langle\sigma_z\rangle_{\rho_t}=\mathrm{Tr}(\rho_t\sigma_z)$ depends on the particular realization of the trajectory and thus on the measurement record $\{V(\tau)\}_{\tau}$ at all times $\tau<t$. The density matrix of the qubit is then reconstructed using the Stochastic Master Equation (SME)~\cite{ficheux2017dynamics}
\begin{equation}
\mathrm{d\rho_t}=-\mathrm{d}t\frac{i}{\hbar}[H,\rho_t]+\mathrm{d}t\frac{\Gamma_m}{2}\mathcal{D}[\sigma_z]\rho_t
+\ \mathrm{d}\mathcal{W}_t\sqrt{2\eta\Gamma_m}\mathcal{M}[\sigma_z]\rho_t,
\label{SME}
\end{equation}
where $\mathcal{D}$ is the Lindblad super operator and $\mathcal{M}$ the measurement super operator $\mathcal{M}[c]\rho=\frac{1}{2}\big((c)-\langle c\rangle)\rho+\rho(c^\dagger-\langle c^\dagger\rangle)\big)$. The two first terms correspond to a Lindbladian evolution of the qubit dephased by the measurement drive (for the seek of simplicity other decoherence channels as spontaneous decay of the qubit have been omitted). The last one represents the measurement backaction: at each $\mathrm{d}t$ the state is kicked depending on the measurement record, possibly changing the mean energy of the qubit. The inherent stochasticity of the SME highlights the profound link between information and energy in quantum mechanics and has triggered recent works on the subject, including in the field of superconducting circuits. The reader can refer to the chapters~\chqtraj,~\chinfera, and~\chthinf for a more precise treatment of the subject.

\end{document}